\documentclass[aps,preprint,floatfix,nofootinbib,showpacs]{revtex4-1}

\pdfoutput=1
\usepackage{graphicx,color,float}
\usepackage{hyperref}
\usepackage{multirow}
\usepackage{verbatim}
\usepackage{longtable}
 \usepackage{amsmath}
\usepackage{amsfonts}
\usepackage{amssymb}
\usepackage{rotating}
\usepackage[dvipsnames]{xcolor}

\usepackage{setspace}

\newcommand{\imag}{\Im {\rm m}}
\newcommand{\real}{\Re {\rm e}}

\newcommand{\gsim}{\raisebox{-0.13cm}{~\shortstack{$>$ \\[-0.07cm] $\sim$}}~}
\def\slash#1{#1\!\!\!/}

\begin{document}

{\small
\begin{flushright}
IUEP-HEP-20-02
\end{flushright} }

\title{
Alignment of Yukawa couplings in two Higgs doublet models
}

\def\slash#1{#1\!\!/}

\renewcommand{\thefootnote}{\arabic{footnote}}

\author{Seong Youl Choi$^{1}$,
Jae Sik Lee$^{2,3,4}$, and
Jubin Park$^{3,4}$} 
 
\affiliation{
$^1$ Department of Physics and RIPC, 
Jeonbuk National University, Jeonju 54896, Korea\\
$^2$ Department of Physics, Chonnam National University,
Gwangju 61186, Korea\\
$^3$ IUEP, Chonnam National University, Gwangju 61186, Korea \\
$^4$ APCTP, Pohang, Gyeongbuk 37673, Korea 
}
\date{November 10, 2020}

\begin{abstract}
\begin{spacing}{1.30}
We study the alignment of Yukawa couplings 
in the framework of general two Higgs doublet models (2HDMs)
considering a scenario in which the lightest neutral 
Higgs boson is purely CP even 
while the two heavier neutral Higgs bosons are allowed to mix 
in the presence of nontrivial CP-violating phases in the Higgs potential.
Identifying the lightest neutral
Higgs boson as the 125 GeV one discovered at the LHC,
we find that the alignment of Yukawa couplings without decoupling
could be easily achieved in the type-I 2HDM  with no much conflict with the current LHC 
Higgs precision data.
Otherwise, we observe that the Yukawa couplings of the lightest Higgs boson
could decouple much slowly
compared to the Higgs coupling to a pair of massive vector bosons and 
they significantly deviate from the corresponding SM values even when the
deviation of the Higgs to vector boson coupling is below the percent level.
On the other hand, independently of 2HDM type and regardless of decoupling,
we find a wrong-sign alignment limit of the Yukawa couplings
in which
the Yukawa couplings to the down-type quarks and/or
those to the charged leptons are equal in strength but
opposite in sign to the corresponding SM ones. 
The magnitude and sign of the up-type quark Yukawa couplings
remain the same as in the SM. Accordingly,
in this limit, all four types of 2HDMs are viable against
the LHC Higgs precision data.
\end{spacing}
\end{abstract}

\maketitle

\section{Introduction} 
Since the discovery of the 125 GeV Higgs boson in 2012 at the
LHC \cite{Aad:2012tfa,Chatrchyan:2012ufa}, it has been scrutinized
very closely and extensively.
At the early stage, several model-independent studies 
\cite{Carmi:2012yp,Azatov:2012bz,Espinosa:2012ir,
Klute:2012pu,Carmi:2012zd,Low:2012rj,Giardino:2012dp,Ellis:2012hz,Espinosa:2012im,
Carmi:2012in,Banerjee:2012xc,Bonnet:2012nm,Plehn:2012iz,Djouadi:2012rh,Dobrescu:2012td,
Cacciapaglia:2012wb,Belanger:2012gc,Moreau:2012da,Corbett:2012dm,Corbett:2012ja,
Masso:2012eq,Cheung:2013kla,Cheung:2014noa}
show that there were some rooms for it to be unlike the one predicted 
in the Standard Model (SM) but, after combining  all the LHC Higgs data 
at 7 and 8 TeV~\cite{Khachatryan:2016vau} and especially those
at 13 TeV~\cite{ATLAS:2018uso,Sirunyan:2018ouh,ATLAS:2018bsg,CMS:2018mmw,
ATLAS:2018gcr,CMS:2018xuk,Aaboud:2018zhk,CMS:2018lkl,Sirunyan:2018kst,ATLAS:2018lur,
CMS:2018nqp,Aaboud:2018urx,Aaboud:2017jvq,Aaboud:2017rss,Sirunyan:2018shy,
Sirunyan:2018ygk,Sirunyan:2018mvw,Sirunyan:2018koj,Aad:2019mbh},
it turns out that it is best described by the SM Higgs boson.
Specifically, the third-generation Yukawa couplings have been established. And
the most recent model-independent study~\cite{Cheung:2018ave} shows that
the $1\sigma$ error of the top-quark Yukawa coupling is
about 6\% while those of the bottom-quark and tau-lepton ones are 
about 10\%. In addition, the possibility of negative top-quark Yukawa coupling
has been completely ruled out and the bottom-quark Yukawa coupling
shows a preference of the positive sign 
\footnote{
Precisely speaking, here the sign of the bottom-quark Yukawa coupling
is relative to the top-quark Yukawa coupling 
configured through the $b$- and $t$-quark loop contributions to 
the $Hgg$ vertex.
}
at about $1.5\sigma$ level.
For the tau-Yukawa coupling, the current data
still do not show any preference for its sign yet.
On the other hand, the coupling to a pair of massive vector bosons
is constrained to be consistent with the SM value
within about 5\% at $1\sigma$ level.

Even though we have not seen any direct hint or evidence of new physics beyond the SM (BSM), 
we are eagerly anticipating it motivated by the tiny but non-vanishing neutrino
masses, matter dominance of our Universe and its evolution driven by dark energy and
dark matters, etc.
In many BSM models, the Higgs sector is extended and it
results in existence of several neutral and
charged Higgs bosons. In this case, one of the neutral Higgs bosons
should be identified as the observed one at the LHC which weighs 125.5 GeV
and its couplings are strongly
constrained to be very SM-like by the current LHC data
as outlined above.
One of the popular ways to achieve this
is to identify the lightest neutral Higgs boson as the 125.5 GeV one
and assume that all the other Higgs bosons are heavier or much heavier than 
the lightest one~\cite{Haber:1989xc,Gunion:2002zf}.
But this decoupling scenario is not phenomenologically interesting and
another scenario is suggested in which all the couplings of the SM-like Higgs candidate 
are (almost) {\it aligned} with those of the SM Higgs while the other Higgs bosons
are not so heavy~\cite{Craig:2013hca,Carena:2013ooa}.

The measure usually taken for decoupling and/or alignment is the Higgs coupling
to a pair of $W$ and $Z$ bosons. In contrast, much attention has not been paid
to the Yukawa couplings.
In this work, we study the decoupling and alignment of Yukawa couplings 
in the framework of the two Higgs doublet models (2HDMs) where the SM is extended by 
adding one more SU(2)$_L$ doublet~\cite{Branco:2011iw}.
Particularly, we consider a scenario 
in which the lightest neutral Higgs boson is purely CP even 
while the two heavier ones might mix 
in the presence of nontrivial CP-violating (CPV) phases in the Higgs potential.

In the type-I 2HDM, we find the Yukawa couplings are well aligned 
even without decoupling as long as $\tan\beta \gsim 2$ and
the 125 GeV Higgs boson discovered at the LHC can be easily accommodated 
with no conflict with the current LHC Higgs precision data.
Otherwise, we find that the Yukawa couplings decouple slowly 
compared to the Higgs coupling to a pair of massive vector bosons
and the alignment with decoupling occurs much slower
when $\tan\beta$ is large.
Interestingly, 
we find a wrong-sign alignment limit of the Yukawa couplings 
regardless of decoupling where
the Yukawa couplings to the down-type quarks in type-II and -IV 2HDMs
and those to the charged leptons in type-II and -III 2HDMs 
take the opposite signs to the corresponding SM ones
while their magnitudes remain the same as in the SM.
Note that both magnitude and sign of the up-type quark 
Yukawa couplings are the same as those of the SM in this limit.
In this limit, all four types of 2HDMs are viable against
the LHC Higgs precision data.

This paper is organized as follows.
Section 2 is devoted to a brief review of the 2HDM Higgs potential, 
the mixing among neutral Higgs bosons and their couplings  to the SM particles.
In Section 3, we build up a scenario in which the lightest Higgs boson is purely CP even
while the two heavier Higgs bosons exhibit CPV mixing.
And we take the scenario to study the alignment of Yukawa couplings in Section 4.
Conclusions are made in Section 5.

\section{Two Higgs Doublet Models}
In this work, we consider the general 2HDMs in which
the potential might be parameterized with three (two real and one complex) massive parameters
and four real and three complex dimensionless quartic couplings as~\cite{Battye:2011jj}:
\begin{eqnarray}
\label{V2HDM}
V_{\rm 2HDM} &=&
+\mu_1^2 (\Phi_1^{\dagger} \Phi_1)
+\mu_2^2 (\Phi_2^{\dagger} \Phi_2)
+m_{12}^2 (\Phi_1^{\dagger} \Phi_2)
+m_{12}^{*2}(\Phi_2^{\dagger} \Phi_1) \nonumber \\
&&+ \lambda_1 (\Phi_1^{\dagger} \Phi_1)^2 + \lambda_2
(\Phi_2^{\dagger} \Phi_2)^2 + \lambda_3 (\Phi_1^{\dagger}
\Phi_1)(\Phi_2^{\dagger} \Phi_2) + \lambda_4 (\Phi_1^{\dagger}
\Phi_2)(\Phi_2^{\dagger} \Phi_1) \nonumber \\
&&+ \lambda_5 (\Phi_1^{\dagger} \Phi_2)^2 +
\lambda_5^{*} (\Phi_2^{\dagger} \Phi_1)^2 + \lambda_6
(\Phi_1^{\dagger} \Phi_1) (\Phi_1^{\dagger} \Phi_2) + \lambda_6^{*}
(\Phi_1^{\dagger} \Phi_1)(\Phi_2^{\dagger} \Phi_1) \nonumber \\
&& + \lambda_7 (\Phi_2^{\dagger} \Phi_2) (\Phi_1^{\dagger} \Phi_2) +
\lambda_7^{*} (\Phi_2^{\dagger} \Phi_2) (\Phi_2^{\dagger} \Phi_1)\; .
\end{eqnarray}
Note that the ${\mathbf Z}_2$ symmetry under $\Phi_1 \to \pm\Phi_1$ and
$\Phi_2 \to \mp\Phi_2$ is hardly broken 
by the no-vanishing quartic couplings $\lambda_6$ and $\lambda_7$ 
and, in this case, we have three rephasing-invariant CPV phases 
in the potential as we will see.
With the parameterization
\begin{equation}
\Phi_1=\left(\begin{array}{c}
\phi_1^+ \\ \frac{1}{\sqrt{2}}\,(v_1+\phi_1+ia_1)
\end{array}\right)\,; \ \ \
\Phi_2={\rm e}^{i\xi}\,\left(\begin{array}{c}
\phi_2^+ \\ \frac{1}{\sqrt{2}}\,(v_2+\phi_2+ia_2)
\end{array}\right)
\end{equation}
and denoting $v_1=v \cos\beta=vc_\beta$
and $v_2=v \sin\beta=vs_\beta$, one may remove $\mu_1^2$, $\mu_2^2$,
and $\imag(m_{12}^2{\rm e}^{i\xi})$ from the 2HDM potential using three tadpole
conditions:
\begin{eqnarray}
\mu_1^2 &=& - v^2\left[\lambda_1c_\beta^2+\frac{1}{2}\lambda_3s_\beta^2+
c_\beta s_\beta\real(\lambda_6{\rm e}^{i\xi})\right]+s_\beta^2 M_{H^\pm}^2\,,
\nonumber \\
\mu_2^2 &=& - v^2\left[\lambda_2s_\beta^2+\frac{1}{2}\lambda_3c_\beta^2+
c_\beta s_\beta\real(\lambda_7{\rm e}^{i\xi})\right]+c_\beta^2 M_{H^\pm}^2\,,
\nonumber \\
\imag(m_{12}^2{\rm e}^{i\xi}) &=& -
\frac{v^2}{2}\left[
2\ c_\beta s_\beta\imag(\lambda_5{\rm e}^{2i\xi})+
c_\beta^2\imag(\lambda_6{\rm e}^{i\xi})+
s_\beta^2\imag(\lambda_7{\rm e}^{i\xi})
\right]\,,
\end{eqnarray}
with the charged Higgs-boson mass  given by
\begin{equation}
M_{H^\pm}^2=
-\frac{\real(m_{12}^2{\rm e}^{i\xi})}{c_\beta s_\beta}
-\frac{v^2}{2c_\beta s_\beta}\left[\lambda_4 c_\beta s_\beta+
2\ c_\beta s_\beta\real(\lambda_5{\rm e}^{2i\xi})+
c_\beta^2\real(\lambda_6{\rm e}^{i\xi})+
s_\beta^2\real(\lambda_7{\rm e}^{i\xi})
\right]\,.
\end{equation}
Then, to fully specify the general 2HDM potential,
one may need the following 13  parameters plus one sign:
\begin{eqnarray}
\label{eq:2hdmpara}
&& v \,, t_\beta\,, |m_{12}| \,;  \nonumber \\
&& \lambda_1\,, \lambda_2\,,\lambda_3\,,\lambda_4\,,
|\lambda_5|\,,|\lambda_6|\,,|\lambda_7|\,; \nonumber \\
&&
\phi_5+2\xi\,,\phi_6+\xi\,,\phi_7+\xi\,,{\rm sign}[\cos(\phi_{12}+\xi)]\,,
\end{eqnarray}
including the vacuum expectation value $v$ and three rephasing-invariant
CPV phases.
Here $m_{12}^2=|m_{12}|^2{\rm e}^{i\phi_{12}}$ and
$\lambda_{5,6,7}=|\lambda_{5,6,7}|{\rm e}^{i\phi_{5,6,7}}$.
We observe that $\sin(\phi_{12}+\xi)$ is fixed by the CP-odd tadpole condition
for $\imag(m_{12}^2{\rm e}^{i\xi})$
when the three CPV phases $\phi_5+2\xi\,,\phi_6+\xi$ and $\phi_7+\xi$ are given
and, accordingly,
$\cos(\phi_{12}+\xi)$ is determined up to the two-fold ambiguity.
One may take the convention with $\xi=0$ without loss of generality.
%


The Higgs potential includes the mass terms which can be cast into the form
\begin{equation}
V_{\rm 2HDM,\,mass}=
M_{H^\pm}^2  H^+H^-
\ + \ \frac{1}{2}
(\phi_1 \ \phi_2  \ a)\,{\cal M}^2_0\,
\left(\begin{array}{c}
\phi_1 \\ \phi_2  \\ a \end{array}\right)
\end{equation}
where $H^\pm$ and $a$ along with the Goldstone bosons of $G^{\pm}$
and $G^0$ are defined through the following rotations:
\begin{equation}
\left(\begin{array}{c} \phi_1^- \\ \phi_2^- \end{array}\right) =
\left(\begin{array}{rr}
c_\beta & -s_\beta \\ s_\beta & c_\beta \end{array}\right)\,
\left(\begin{array}{c} G^- \\ H^- \end{array}\right) \,; \ \ \
\left(\begin{array}{c} a_1 \\ a_2 \end{array}\right) =
\left(\begin{array}{rr}
c_\beta & -s_\beta \\ s_\beta & c_\beta \end{array}\right)\,
\left(\begin{array}{c} G^0 \\ a \end{array}\right) \,.
\end{equation}
And the $3\times 3$ mass matrix of the neutral Higgs bosons
${\cal M}_0^2$ is given by
\begin{equation}
\label{eq:2hdmm0sq}
{\cal M}^2_0 = M_A^2 \left(\begin{array}{ccc}
s_\beta^2 & -s_\beta c_\beta & 0 \\
-s_\beta c_\beta & c_\beta^2 & 0 \\
0 & 0 & 1 \end{array}\right) \ + \ {\cal M}^2_\lambda
\end{equation}
with 
\begin{eqnarray}
M_A^2&=&M_{H^\pm}^2+\frac{1}{2}\lambda_4 v^2 -\real(\lambda_5{\rm
e}^{2i\xi})v^2\,,
\end{eqnarray}
and
\begin{equation}
\frac{{\cal M}^2_\lambda}{v^2} = \left(\begin{array}{lll}
2\lambda_1 c_\beta^2 +2\real(\lambda_5{\rm e}^{2i\xi})s_\beta^2 &
\lambda_{34}c_\beta s_\beta + \real(\lambda_6{\rm e}^{i\xi}) c_\beta^2 &
-\imag(\lambda_5{\rm e}^{2i\xi})s_\beta \\
+2\real(\lambda_6{\rm e}^{i\xi}) s_\beta c_\beta &
+\real(\lambda_7{\rm e}^{i\xi}) s_\beta^2 &
-\imag(\lambda_6{\rm e}^{i\xi}) c_\beta \\[5mm]
\lambda_{34}c_\beta s_\beta + \real(\lambda_6{\rm e}^{i\xi}) c_\beta^2 &
2\lambda_2 s_\beta^2 +2\real(\lambda_5{\rm e}^{2i\xi})c_\beta^2 &
-\imag(\lambda_5{\rm e}^{2i\xi})c_\beta \\
+\real(\lambda_7{\rm e}^{i\xi}) s_\beta^2 &
+2\real(\lambda_7{\rm e}^{i\xi}) s_\beta c_\beta &
-\imag(\lambda_7{\rm e}^{i\xi}) s_\beta \\[3mm]
-\imag(\lambda_5{\rm e}^{2i\xi})s_\beta &
-\imag(\lambda_5{\rm e}^{2i\xi})c_\beta &
~~~~~~~~~0 \\
-\imag(\lambda_6{\rm e}^{i\xi}) c_\beta &
-\imag(\lambda_7{\rm e}^{i\xi}) s_\beta &
\end{array}\right)\,.
\end{equation}
where $\lambda_{34}=\lambda_3+\lambda_4$.
%
%
%


Once the mass matrix is given,
the $3\times 3$ mixing matrix $O$ is defined through
\begin{eqnarray}
(\phi_1,\phi_2,a)^T_\alpha&=&O_{\alpha i} (H_1,H_2,H_3)^T_i
\end{eqnarray}
such that $O^T {\cal M}_0^2 O={\rm diag}(M_{H_1}^2,M_{H_2}^2,M_{H_3}^2)$
with the ordering of $M_{H_1}\leq M_{H_2}\leq M_{H_3}$.


The trilinear interactions of the neutral and charged Higgs bosons
with the gauge bosons $Z$ and $W^\pm$
are described by the three interaction Lagrangians:
\begin{eqnarray}
\label{eq:hvvCouplings}
{\cal L}_{HVV} & = & g\,M_W \, \left(W^+_\mu W^{- \mu}\ + \
\frac{1}{2c_W^2}\,Z_\mu Z^\mu\right) \, \sum_i \,g_{_{H_iVV}}\, H_i
\,,\\[3mm]
{\cal L}_{HHZ} &=& \frac{g}{2c_W} \sum_{i>j} g_{_{H_iH_jZ}}\, Z^{\mu}
(H_i\, \!\stackrel {\leftrightarrow} {\partial}_\mu H_j) \,, \\ [3mm]
{\cal L}_{HH^\pm W^\mp} &=& -\frac{g}{2} \, \sum_i \, g_{_{H_iH^+
W^-}}\, W^{-\mu} (H_i\, i\!\stackrel{\leftrightarrow}{\partial}_\mu
H^+)\, +\, {\rm h.c.}\,,
\end{eqnarray}
where $i,j =1,2,3$ and
the couplings $g_{_{H_iVV}}$, $g_{_{H_iH_jZ}}$ and $g_{_{H_iH^+
W^-}}$ are given in terms of the neutral Higgs-boson mixing matrix $O$
by (note that det$(O)=\pm1$ for any orthogonal matrix $O$):
\begin{eqnarray}
g_{_{H_iVV}} &=& c_\beta\, O_{\phi_1 i}\: +\: s_\beta\, O_{\phi_2 i}
\, ,\nonumber \\
g_{_{H_iH_jZ}} &=& {\rm sign} [{\rm det}(O)] \, \, \varepsilon_{ijk}\,
g_{_{H_kVV}}\,,  \nonumber \\
g_{_{H_iH^+ W^-}} &=& c_\beta\, O_{\phi_2 i} - s_\beta\, O_{\phi_1 i}
- i O_{ai} \, ,
\end{eqnarray}
leading to the following sum rules:
\begin{equation}
\sum_{i=1}^3\, g_{_{H_iVV}}^2\ =\ 1\,\quad{\rm and}\quad
g_{_{H_iVV}}^2+|g_{_{H_iH^+ W^-}}|^2\ =\ 1\,\quad {\rm for~ each}~ i\,.
\end{equation}

On the other hand, the Yukawa couplings in 2HDMs could be written as~\cite{Cheung:2013rva}:
\begin{eqnarray}
-{\cal L}_Y&=& h_u\, \overline{u_R}\, Q^T\,(i\tau_2)\,\Phi_2
-h_d\, \overline{d_R}\, Q^T\,(i\tau_2)\,
\left(\eta_1^d\,\widetilde\Phi_1 +\eta_2^d\,\widetilde\Phi_2\right)
\nonumber \\[2mm] &-&
h_l\, \overline{l_R}\, L^T\,(i\tau_2)\,
\left(\eta_1^l\,\widetilde\Phi_1 +\eta_2^l\,\widetilde\Phi_2\right)
\ + \ {\rm h.c.}
\end{eqnarray}
where $Q^T=(u_L\,,d_L)$, $L^T=(\nu_L\,,l_L)$, and
$\widetilde\Phi_i=i\tau_2 \Phi_i^*$.
Note that, in our convention,  we make
the up-type quarks coupled to only $\Phi_2$ exploiting a freedom to
redefine the two linear combinations of $\Phi_1$ and $\Phi_2$~\cite{Davidson:2005cw}.
And, the down-type quarks and the charged leptons are coupled to 
either $\Phi_1$ or $\Phi_2$ in ${\mathbf Z}_2$ symmetric way
leading to the 4 types of 2HDMs without tree-level flavor-changing neutral currents (FCNCs)
as classified in Table \ref{tab:2hdtype}.
%
\begin{table}[!t]
\caption{\label{tab:2hdtype}
Classification of 2HDMs satisfying the Glashow-Weinberg condition
\cite{Glashow:1976nt} which guarantees the absence of tree-level FCNC.}
\begin{center}
\begin{tabular}{|l|cccc|}
\hline
\hline
         & \hspace{0.5cm} 2HDM I\hspace{0.5cm} & 2HDM II\hspace{0.5cm}
& 2HDM III\hspace{0.5cm} & 2HDM IV\hspace{0.5cm}  \\
\hline
$\eta_1^d$  & $0$ & $1$ & $0$ & $1$  \\
$\eta_2^d$  & $1$ & $0$ & $1$ & $0$  \\
\hline
\hline
$\eta_1^l$  & $0$ & $1$ & $1$ & $0$  \\
$\eta_2^l$  & $1$ & $0$ & $0$ & $1$  \\
\hline
\hline
\end{tabular}
\end{center}
\end{table}
By identifying the couplings
\footnote{
Here we take the convention with $\xi=0$ and the couplings $h_{u,d,l}$ are
supposed to be real.}
\begin{equation}
h_u = \frac{\sqrt{2}m_u}{v}\,\frac{1}{s_\beta}\,; \ \
h_d = \frac{\sqrt{2}m_d}{v}\,\frac{1}{\eta_1^dc_\beta+\eta_2^d s_\beta}\,; \ \
h_l = \frac{\sqrt{2}m_l}{v}\,\frac{1}{\eta_1^lc_\beta+\eta_2^l s_\beta}\,,
\end{equation}
the couplings of three neutral Higgs bosons to a pair of fermions 
could be cast into the form collectively:
\begin{eqnarray}
\label{eq:nhff.2hdm}
{\cal L}_{H_i\bar{f}f} &=&-\sum_{f=u,d,\ell} \frac{m_f}{v}\,
\sum_{i=1}^3\left[ \bar{f}\, \left(
g^S_{H_i\bar f f} +i\,\gamma_5\,g^P_{H_i\bar f f}
\right)\,f \right]\,H_i\,.
\end{eqnarray}
For the SM Higgs, the normalized Yukawa couplings are given by
$g^S_{H_{\rm SM}\bar f f}=1$ and $g^P_{H_{\rm SM}\bar f f}=0$.


%
\begin{table}[!t]
\caption{The 
normalized Yukawa couplings of the neutral Higgs bosons of $H_{i}$ ($i=1,2,3$) to
a pair of fermions in the four types of 2HDMs in terms of the elements of
the mixing matrix $O$ and $t_\beta$, see Eq.~(\ref{eq:nhff.2hdm}.}
\label{tab:2hdmCouplings}
\setlength{\tabcolsep}{1.5ex}%
\begin{center}
\vspace{-2mm}
\begin{tabular}{c|rrrr}
\hline\hline
 &  I~~~~ & II~~~~ & III~~~~ & IV~~~~ \\
\hline
$g^S_{H_i\bar t t\,,H_i\bar c c\,,H_i\bar u u}$ &
$O_{\phi_2i}/s_\beta$ & $O_{\phi_2i}/s_\beta$ &
$O_{\phi_2i}/s_\beta$ & $O_{\phi_2i}/s_\beta$ \\
$g^S_{H_i\bar b b\,,H_i\bar s s\,,H_i\bar d d}$ &
$O_{\phi_2i}/s_\beta$ & $O_{\phi_1i}/c_\beta$ &
$O_{\phi_2i}/s_\beta$ & $O_{\phi_1i}/c_\beta$ \\
$g^S_{H_i\tau \tau\,,H_i\mu\mu\,,H_i ee}$ &
$O_{\phi_2i}/s_\beta$ & $O_{\phi_1i}/c_\beta$ &
$O_{\phi_1i}/c_\beta$ & $O_{\phi_2i}/s_\beta$ \\
\hline
$g^P_{H_i\bar t t\,,H_i\bar c c\,,H_i\bar u u}$ &
$-O_{ai}/t_\beta$ & $-O_{ai}/t_\beta$ &
$-O_{ai}/t_\beta$ & $-O_{ai}/t_\beta$ \\
$g^P_{H_i\bar b b\,,H_i\bar s s\,,H_i\bar d d}$ &
$O_{ai}/t_\beta$ & $-O_{ai}\, t_\beta$ &
$O_{ai}/t_\beta$ & $-O_{ai}\, t_\beta$ \\
$g^P_{H_i\tau \tau\,,H_i\mu\mu\,,H_i ee}$ &
$O_{ai}/t_\beta$ &  $-O_{ai}\,t_\beta$ &
$-O_{ai}\,t_\beta$ & $O_{ai}/t_\beta$ \\
\hline\hline
\end{tabular}
\end{center}
\end{table}
We note all the neutral Higgs-boson couplings to fermions 
and massive vector bosons are fully determined once
the mixing matrix $O$ and $t_\beta$ are given, see Table \ref{tab:2hdmCouplings}.
and Eq.~(\ref{eq:hvvCouplings}).

\section{A scenario}
In this section, we study the mixing in the neutral Higgs boson sector
of 2HDMs in the presence of nontrivial CPV phases in the
quartic couplings of $\lambda_{5,6,7}$.
To be specific, we consider a scenario in which
the lightest neutral Higgs boson is purely CP even while
a CPV mixing might occur between the two heavy neutral Higgs bosons 
which are mixtures of CP-even and CP-odd states.  
When a CPV mixing in the neutral heavy Higgs sector occurs, the couplings of
the two neutral heavy Higgs bosons to a pair of massive gauge bosons 
exist simultaneously.
Note that, in the CP-conserving (CPC) case,
one of the neutral heavy Higgs 
bosons is purely CP odd and its coupling
to a pair of massive gauges bosons is identically vanishing.


For general  CPV scenarios in 2HDMs,  all of the three neutral Higgs bosons
do not carry definite CP parities and they become mixtures of CP-even and CP-odd
states.  In this case, without loss of generality,
the orthogonal $3\times 3$ mixing matrix $O$ might be parameterized as
\begin{eqnarray}
\label{eq:omix}
\hspace{-0.5cm}
\vspace{ 1.0cm}
O &=&
\left( \begin{array}{ccc}
 -s_\alpha  &   c_\alpha   &  0   \\
  c_\alpha  &   s_\alpha   &  0   \\
  0         &      0       &  1   \\
  \end{array} \right)
\left( \begin{array}{ccc}
    c_\eta               &      0             &   s_\eta  \\
    0                    &      1             &   0   \\
   -s_\eta               &      0             &   c_\eta   \\
  \end{array} \right) 
\left( \begin{array}{ccc}
    1               &      0       &   0   \\
    0               &   c_\omega   &   s_\omega   \\
    0               &  -s_\omega   &   c_\omega   \\
  \end{array} \right)  
\nonumber \\[5mm] &=&
\left( \begin{array}{ccc}
    -s_\alpha c_\eta  &  c_\alpha c_\omega + s_\alpha s_\eta s_\omega   &
    c_\alpha s_\omega - s_\alpha s_\eta c_\omega     \\
    c_\alpha c_\eta   &  s_\alpha c_\omega - c_\alpha s_\eta s_\omega   &
    s_\alpha s_\omega + c_\alpha s_\eta c_\omega     \\
    -s_\eta  &  -c_\eta s_\omega   &    c_\eta c_\omega  \\
  \end{array} \right)\,, 
\end{eqnarray}
introducing a CP-conserving (CPC) mixing angle $\alpha$ and two
CPV angles $\omega$ and $\eta$ which should be fixed to study
Higgs decays in the 2HDM framework.
Note that, in our parameterization of the mixing matrix $O$,  we have
${\rm det}(O)=-1$ to follow the CPC convention usually taken in the literature:
\begin{equation}
\left( \begin{array}{c} H \\ h \end{array} \right) =
\left( \begin{array}{cc}
  c_\alpha  &   s_\alpha \\
 -s_\alpha  &   c_\alpha
  \end{array} \right)
\left( \begin{array}{c} \phi_1 \\ \phi_2 \end{array} \right) \,,
\end{equation}
in terms of the two CP-even states of $H$ (heavier) and $h$ (lighter).
%


We take $s_\eta=0$ and $c_\eta=+1$ assuming that
the lightest Higgs boson $H_1$ is purely CP even and, in this case, the mixing
matrix $O$ takes the form of
\begin{equation}
O \ = \
\left( \begin{array}{ccc}
   -s_\alpha  &  c_\alpha c_\omega  & c_\alpha s_\omega   \\
  ~~c_\alpha  &  s_\alpha c_\omega  & s_\alpha s_\omega   \\
   ~0  &  -s_\omega   &    c_\omega   \\
  \end{array} \right)\,. 
\end{equation}
We observe $H_3$ is CP odd when $|c_\omega|=1$ while
$H_2$ is CP odd when $|s_\omega|=1$.
With the above $O$, the couplings of three neutral Higgs bosons to a pair of massive
vector bosons are given by
\begin{eqnarray}
g_{_{H_1VV}}  =  s_{\beta-\alpha} \equiv +\sqrt{1-\epsilon}\,, \ \ \
g_{_{H_2VV}}  =  c_{\beta-\alpha} c_\omega \equiv \delta_2\,, \ \ \
g_{_{H_3VV}}  =  c_{\beta-\alpha} s_\omega \equiv \delta_3\,,
\end{eqnarray}
with $\delta_2^2+\delta_3^2=\epsilon$.
Note that we are taking $g_{_{H_1VV}}>0$ and the two mixing angles
$\alpha$ and $\omega$ are determined by giving $\delta_{2,3}$ and $t_\beta$:
\begin{eqnarray}
\label{eq:2hdmaw}
s_\alpha &=& -\sqrt{1-\epsilon}\, c_\beta +\frac{\delta_2}{c_\omega}\, s_\beta 
= -\sqrt{1-\epsilon}\, c_\beta + 
{\rm sign}\left[\frac{\delta_2}{c_\omega}\right]\,\sqrt\epsilon \, s_\beta \,, 
\nonumber \\[2mm]
c_\alpha &=& +\sqrt{1-\epsilon}\, s_\beta +\frac{\delta_2}{c_\omega}\, c_\beta 
=\sqrt{1-\epsilon}\, s_\beta +
{\rm sign}\left[\frac{\delta_2}{c_\omega}\right]\,\sqrt\epsilon \, c_\beta \,,  
\nonumber \\[2mm]
c_\omega^2 &=& \frac{\delta_2^2}{\delta_2^2+\delta_3^2} = 
\frac{\delta_2^2}{\epsilon}\,, \ \ \
s_\omega^2  =  \frac{\delta_3^2}{\delta_2^2+\delta_3^2} =\frac{\delta_3^2}{\epsilon}\,.
\end{eqnarray}


With our choice of  $s_\eta=0$, the relation
$O^T {\cal M}_0^2 O={\rm diag}(M_{H_1}^2,M_{H_2}^2,M_{H_3}^2)$ gives
\begin{eqnarray}
\label{eq:2hdmm0sqij}
\left({\cal M}_0^2\right)_{11} &=&
s_\alpha^2\,M_{H_1}^2+c_\alpha^2 c_\omega^2\,M_{H_2}^2+ c_\alpha^2
s_\omega^2\,M_{H_3}^2\,,
\nonumber \\[2mm]
\left({\cal M}_0^2\right)_{22} &=&
c_\alpha^2\,M_{H_1}^2+s_\alpha^2 c_\omega^2\,M_{H_2}^2+ s_\alpha^2
s_\omega^2\,M_{H_3}^2\,,
\nonumber \\[2mm]
\left({\cal M}_0^2\right)_{33} &=&
s_\omega^2\,M_{H_2}^2+ c_\omega^2\,M_{H_3}^2 = M_A^2\,;
\nonumber \\[2mm]
\left({\cal M}_0^2\right)_{12} &=&
c_\alpha s_\alpha\,\left(-M_{H_1}^2+c_\omega^2\,M_{H_2}^2
+ s_\omega^2\,M_{H_3}^2\right)\,, \nonumber \\[2mm]
\left({\cal M}_0^2\right)_{13} &=&
c_\alpha c_\omega s_\omega\,\left(M_{H_3}^2-M_{H_2}^2\right)\,,
\nonumber \\[2mm]
\left({\cal M}_0^2\right)_{23} &=&
s_\alpha c_\omega s_\omega\,\left(M_{H_3}^2-M_{H_2}^2\right)\,.
\end{eqnarray}
Using the third and fourth relations in the above and Eq.~(\ref{eq:2hdmm0sq}), we have
\begin{eqnarray}
s_\alpha c_\alpha = \frac{\left({\cal M}_0^2\right)_{12}}
{-M_{H_1}^2+c_\omega^2\,M_{H_2}^2 + s_\omega^2\,M_{H_3}^2}
=\frac{-s_\beta c_\beta (s_\omega^2\,M_{H_2}^2+ c_\omega^2\,M_{H_3}^2) 
+ \left({\cal M}_\lambda^2\right)_{12}}
{-M_{H_1}^2+c_\omega^2\,M_{H_2}^2 + s_\omega^2\,M_{H_3}^2}
\end{eqnarray}
which, using Eq.~(\ref{eq:2hdmaw}), turns into
\begin{eqnarray}
\epsilon\,s_{2\beta}-\sqrt\epsilon\sqrt{1-\epsilon}\,c_{2\beta}
=\frac{-c_\beta s_\beta\left[
M_{H_1}^2+c_{2\omega}( M_{H_3}^2 - M_{H_2}^2)\right]+\left({\cal
M}_\lambda^2\right)_{12}}
{-M_{H_1}^2+c_\omega^2 M_{H_2}^2 +s_\omega^2 M_{H_3}^2}\equiv \chi\,.
\end{eqnarray}
By solving it for $\epsilon$, we have
\begin{equation}
\epsilon=\left(\chi\, s_{2\beta}+\frac{c_{2\beta}^2}{2}\right)
-\frac{|c_{2\beta}|}{2}\left[1-\left(2\chi-s_{2\beta}\right)^2\right]^{1/2}
=\frac{1}{c_{2\beta}^2}\,\chi^2 \ - \ \frac{2s_{2\beta}}{c_{2\beta}^4}\,\chi^3
+\cdots\,.
\end{equation}
Note that $2\chi-s_{2\beta}=s_{2\alpha}$.
We further note that $\epsilon$ is suppressed by the quartic powers of the heavy
Higgs-boson masses in the leading order.


In the decoupling limit of
$M_{H_{2,3}}^2 \gg 
M_{H_1}^2 \sim \left({\cal M}_\lambda^2\right)_{12} \sim (M_{H_3}^2 - M_{H_2}^2)$,
$\epsilon$ approaches 0 and, accordingly, one has
\begin{equation}
\delta_{2,3} \to 0\,; \ \ \
s_\alpha \to -c_\beta \,, \ \ \
c_\alpha \to +s_\beta \,. 
\end{equation}
We also observe that, the fifth and sixth relations in Eq.~(\ref{eq:2hdmm0sqij})
lead to
\begin{equation}
t_\alpha =\frac{\left({\cal M}_0^2\right)_{23}}{\left({\cal M}_0^2\right)_{13}}=
\frac{\imag(\lambda_5)+\imag(\lambda_7)t_\beta}
{\imag(\lambda_5)t_\beta + \imag(\lambda_6)} \ \to \ -\frac{1}{t_\beta}
\end{equation}
which gives
\begin{equation}
\imag(\lambda_5) \  \to \ -\frac{1}{2}\left[
\frac{\imag(\lambda_6)}{t_\beta} + \imag(\lambda_7)\, t_\beta \right].
\end{equation}


To summarize, among the 13 free parameters of 2HDMs
counted as in Eq.~(\ref{eq:2hdmpara}), we have addressed 8 parameters of
\begin{equation}
v\,, \ \ t_\beta\,; \ \ s_\eta\,, \ \ \delta_2\,, \ \ \delta_3 ; \ \
M_{H_1}\,, \ \ M_{H_2}\,,\ \ M_{H_3}\,.
\end{equation}
The choice of $s_\eta=0$ makes the lightest Higgs boson purely CP even and
$\delta_2$ and $\delta_3$ ($\epsilon=\delta_2^2+\delta_3^2$)
fix the  two remaining mixing angles of
$\alpha$ and $\omega$ up to signs
with the given value of $t_\beta$,
see Eq.~(\ref{eq:2hdmaw}).
Note that the masses of three neutral Higgs bosons are independent from each
other when $\lambda_6$ and/or $\lambda_7$ are non-vanishing which 
is consistent with results presented in 
Refs.~\cite{ElKaffas:2007rq,Grzadkowski:2014ada,Grzadkowski:2018ohf}.

\section{Alignment of Yukawa couplings}
In this section, we closely examine the couplings of the lightest Higgs 
boson $H_1$ in the scenario represented by Eq.~(\ref{eq:2hdmaw}) 
interpreting the CP-even $H_1$ is the Higgs boson discovered at the LHC with
$M_{H_1}=M_{H_{\rm SM}}=125.5$ GeV. 
In this case, $g^S_{H_1\bar f f}$ and $g_{_{H_1VV}}$
should be close to the SM values of $1$
for the decay pattern of $H_1$ to be consistent with the current LHC Higgs 
precision data.


\begin{figure}[t!]
\centering
\includegraphics[height=5.6in,angle=0]{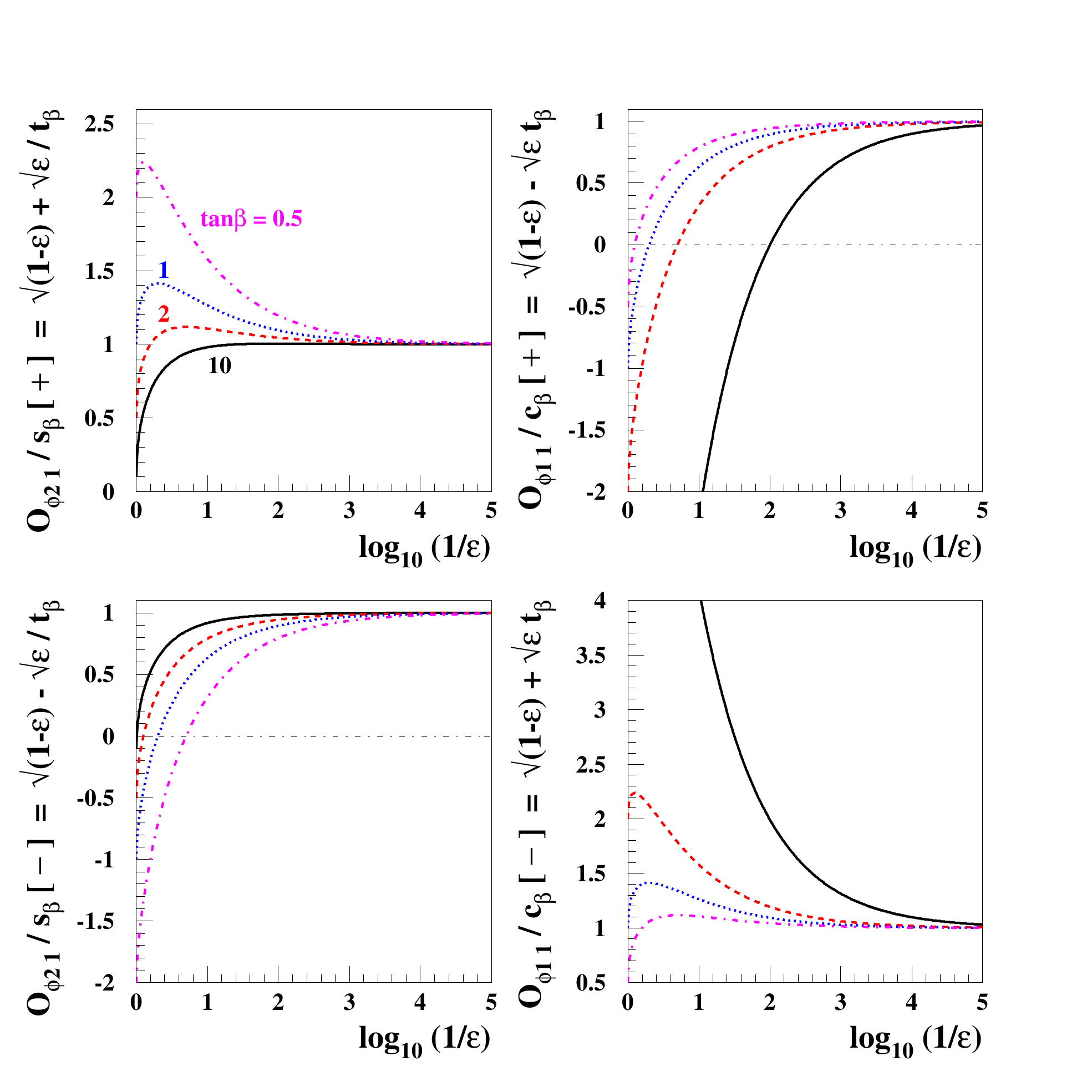}
\caption{\label{fig:gsh1eps}
The normalized Yukawa couplings of $H_1$ to a pair of SM fermions
$g^S_{H_1\bar f f}$
for ${\rm sign}\left[\frac{\delta_2}{c_\omega}\right]=+1$ (upper) and
${\rm sign}\left[\frac{\delta_2}{c_\omega}\right]=-1$ (lower)
as functions of $\log_{10}(1/\epsilon)$ taking
four values of $t_\beta=0.5$ (dash-dotted), $1$ (dotted), $2$ (dashed),
and $10$ (solid).
}
\end{figure}
From Table \ref{tab:2hdmCouplings}, irrelevant of the type of 2HDMs,
we observe that  any $H_1$ coupling to a pair of SM fermions is given by one of
the following two quantities
\footnote{Note that, in our convention, 
both  the lightest neutral Higgs boson couplings
to a pair of vector bosons $V=W,Z$ and its Yukawa couplings to the 
up-type quarks are positive definite
when ${\rm sign}[\delta_2/c_\omega]=+1$ independently of 2HDM type and $t_\beta$.}:
\begin{equation}
\frac{O_{\phi_2 1}}{s_\beta}=\frac{c_\alpha}{s_\beta}=
\sqrt{1-\epsilon}+{\rm sign}\left[\frac{\delta_2}{c_\omega}\right]\frac{\sqrt{\epsilon}}{t_\beta}\,, \ \ \
\frac{O_{\phi_1 1}}{c_\beta}=\frac{-s_\alpha}{c_\beta}=
\sqrt{1-\epsilon}-{\rm sign}\left[\frac{\delta_2}{c_\omega}\right]{\sqrt{\epsilon}}\,{t_\beta}\,,
\end{equation}
with $g^P_{H_1\bar f f}=0$ taking $O_{a1}=-s_\eta=0$.
In Fig.~\ref{fig:gsh1eps}, we show ${O_{\phi_2 1}}/{s_\beta}$ (left) and
${O_{\phi_1 1}}/{c_\beta}$ (right)  as functions of $\log_{10}(1/\epsilon)$
for positive and negative $\delta_2/c_w$ in the upper and lower panels, respectively,
taking four values of $t_\beta$.
First of all, we see that both of the normalized Yukawa couplings approach to or align with
the SM value of $1$ in the decoupling limit where $\epsilon\to 0$.
Otherwise, the couplings are not always positive and their magnitudes
could be significantly larger than 1. Moreover, 
compared to the $g_{_{H_1VV}}=\sqrt{1-\epsilon}$ coupling, they decouple
slowly or much slowly depending on $t_\beta$.
If $\epsilon=0.01$ and $t_\beta =10$, for example, 
the $g_{_{H_1VV}}$ coupling deviates from its SM
value by only 0.5\% but the Yukawa coupling ${O_{\phi_1 1}}/{c_\beta}$
could be as large as 2 or even could vanish 
depending on the sign of ${\delta_2}/{c_\omega}$ resulting 
in 100\% deviation from its SM value: see the black lines
in the upper- and lower-right panels of Fig.~\ref{fig:gsh1eps}.
The similar observation could be made for ${O_{\phi_2 1}}/{s_\beta}$
for small $t_\beta<1$.


%
\begin{figure}[t!]
\centering
\includegraphics[height=4.3in,angle=0]{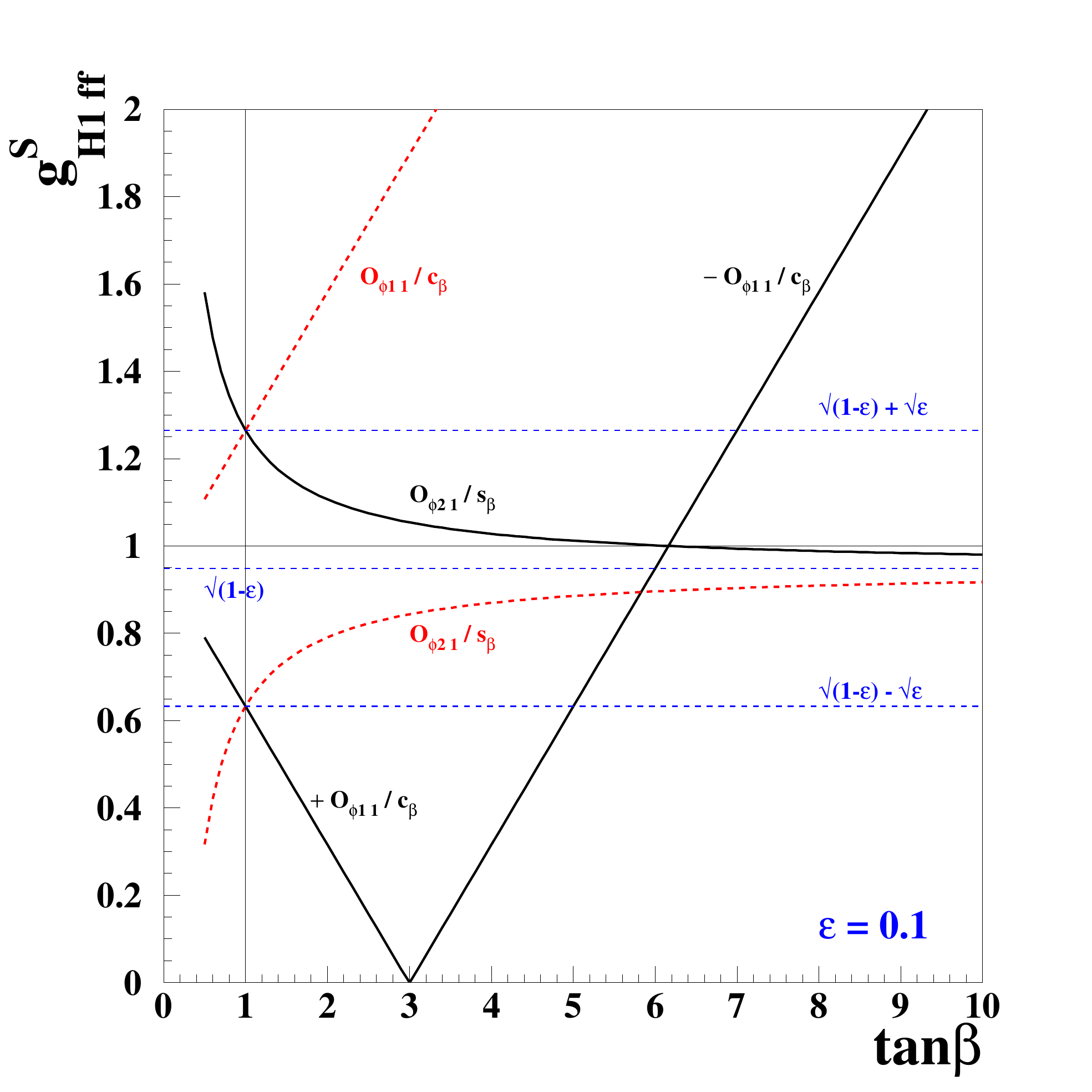}
\caption{\label{fig:gsh1ff}
The normalized Yukawa couplings of $H_1$ to a pair of SM fermions
$g^S_{H_1\bar f f}$
for ${\rm sign}\left[\frac{\delta_2}{c_\omega}\right]=+1$ (black solid) and
${\rm sign}\left[\frac{\delta_2}{c_\omega}\right]=-1$ (red dashed)
as functions of $t_\beta$ taking $\epsilon=0.1$.
Three dashed horizontal blue lines locate the positions of
$\sqrt{1-\epsilon}$ and $\sqrt{1-\epsilon}\pm\sqrt\epsilon$
while the horizontal solid one is for the SM value of 1.
The vertical solid line is for $t_\beta=1$.
}
\end{figure}
In Fig.~\ref{fig:gsh1ff}, we show the couplings of $H_1$ to a pair of the SM fermions
$g^S_{H_1\bar f f}$
as functions of $t_\beta$ when
${\delta_2}/{c_\omega}>0$ (black solid) and
${\delta_2}/{c_\omega}<0$ (red dashed) taking $\epsilon=0.1$.
The coupling $|{O_{\phi_1 1}}/{c_\beta}|$ linearly grows as $t_\beta$ increases
implying that the $H_1$ couplings to bottom quarks in type-II and -IV 2HDMs
and those to tau leptons in type-II and -III 2HDMs could significantly deviate from
the SM value of 1 for intermediate and/or large values of $t_\beta$
depending on the size of $\epsilon$.
In contrast, in the type-I 2HDM, all the Yukawa couplings are given by
$O_{\phi_2 1}/s_\beta$ and it remains more or less aligned with the SM value
as long as $t_\beta \gsim 1$.
%


%
\begin{table}[!t]
\caption{
The normalized Yukawa couplings of the lightest neutral Higgs boson $H_{1}$ 
in the four types of 2HDMs  in the wrong-sign alignment limit where
$t_\beta=(1+\sqrt{1-\epsilon})/\sqrt\epsilon$.}
\label{tab:WSYukawa}
\setlength{\tabcolsep}{1.5ex}%
\begin{center}
\vspace{-2mm}
\begin{tabular}{c|rrrr}
\hline\hline
 &  I & II & III & IV \\
\hline
$g^S_{H_1\bar t t\,,H_1\bar c c\,,H_1\bar u u}$ &
$1$ & $1$ & $1$ & $1$ \\
$g^S_{H_1\bar b b\,,H_1\bar s s\,,H_1\bar d d}$ &
$1$ & $-1$ & $1$ & $-1$ \\
$g^S_{H_1\tau \tau\,,H_1\mu\mu\,,H_1 ee}$ &
$1$ & $-1$ & $-1$ & $1$ \\
\hline\hline
\end{tabular}
\end{center}
\end{table}

When ${\delta_2}/{c_\omega}>0$,
we note there is another possibility in which ${O_{\phi_1 1}}/{c_\beta}$
takes the opposite sign to the SM coupling but its absolute value is $1$~\cite{Gunion:2002zf}.
This occurs at $t_\beta=(1+\sqrt{1-\epsilon})/\sqrt{\epsilon}$ and, surprisingly, 
this also makes ${O_{\phi_2 1}}/{s_\beta}=+1$, see the black solid lines in 
Fig.~\ref{fig:gsh1ff} around $t_\beta = 6$ and Table~\ref{tab:WSYukawa}.
Therefore, every 2HDM could be made consistent with
the LHC Higgs precision data independently of its type
by making this specific choice of $t_\beta$ at the expense
of {\it wrong} signs of the $H_1$ couplings to 
down-type quarks in type-IV 2HDM,
those to charged leptons in type-III 2HDM, and
those to both down-type quarks and charged leptons in type-II 2HDM.
Note that this wrong-sign alignment of the Yukawa couplings
happens for any value of $\epsilon$, whether in the decoupling limit or not.


\begin{figure}[t!]
\centering
\includegraphics[height=3.6in,width=5in]{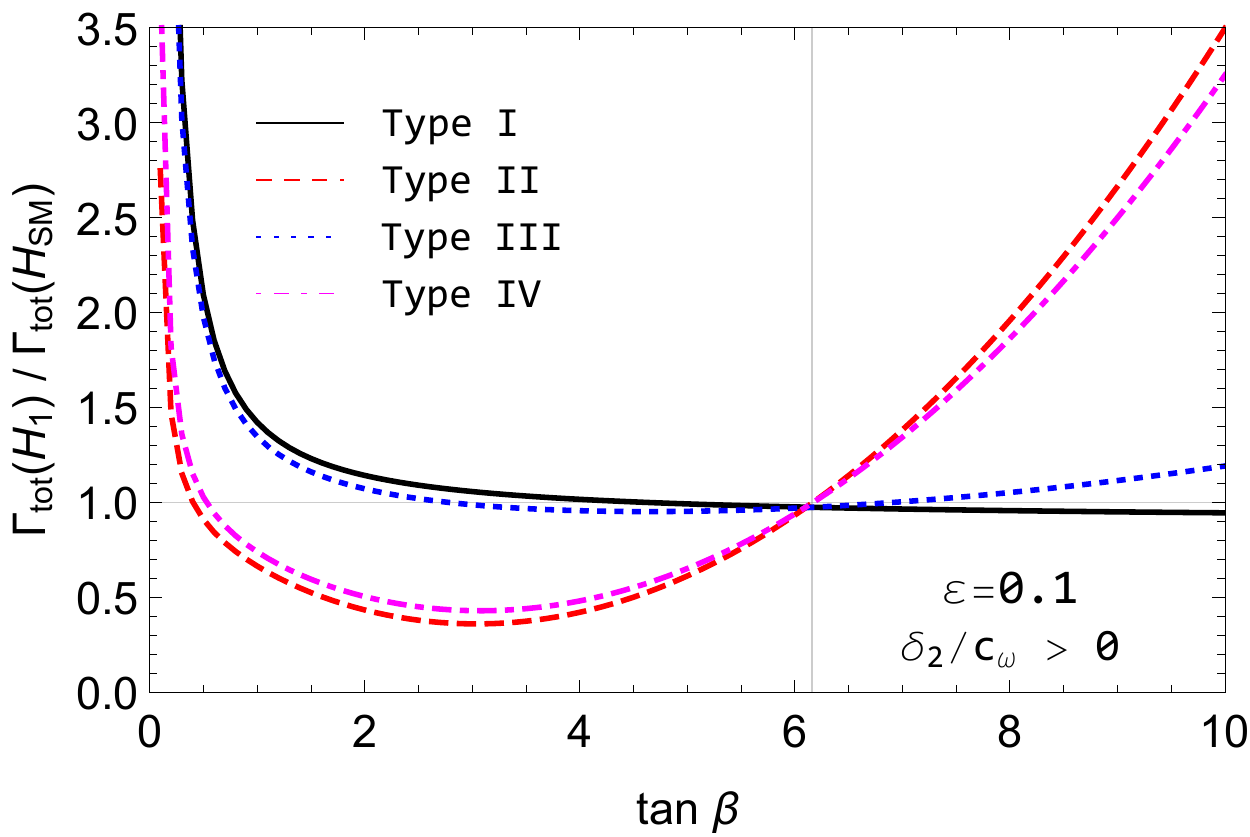}
\caption{\label{fig:h1gam}
The ratio of the total decay width of the lightest Higgs boson $H_1$
to that of the SM,
$\Gamma_{\rm tot}(H_1)/\Gamma_{\rm tot}(H_{\rm SM})$, 
as functions of $t_\beta$
in the four types of 2HDMs: type I   (black solid), type II  (red dashed),
type III (blue dotted), and type IV  (magenta dash-dotted).
We have taken $\epsilon=0.1$ and ${\rm sign}[\delta_2/c_\omega]=+1$.
The SM decay width $\Gamma_{\rm tot}(H_{\rm SM})=4.122$ MeV
with $M_{H_1}=M_{H_{\rm SM}}=125.5$ GeV.}
\end{figure}
\begin{figure}[t!]
\centering
\includegraphics[height=1.6in,width=2in]{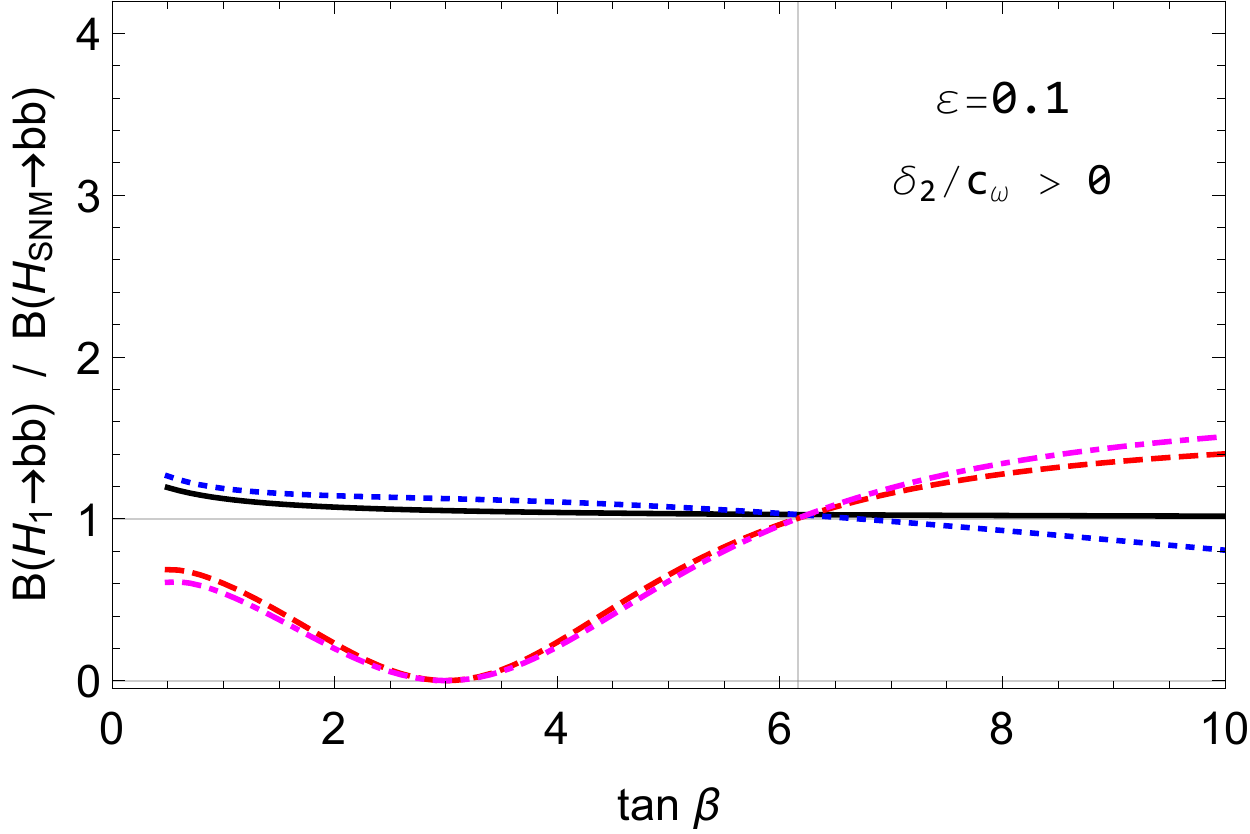}
\includegraphics[height=1.6in,width=2in]{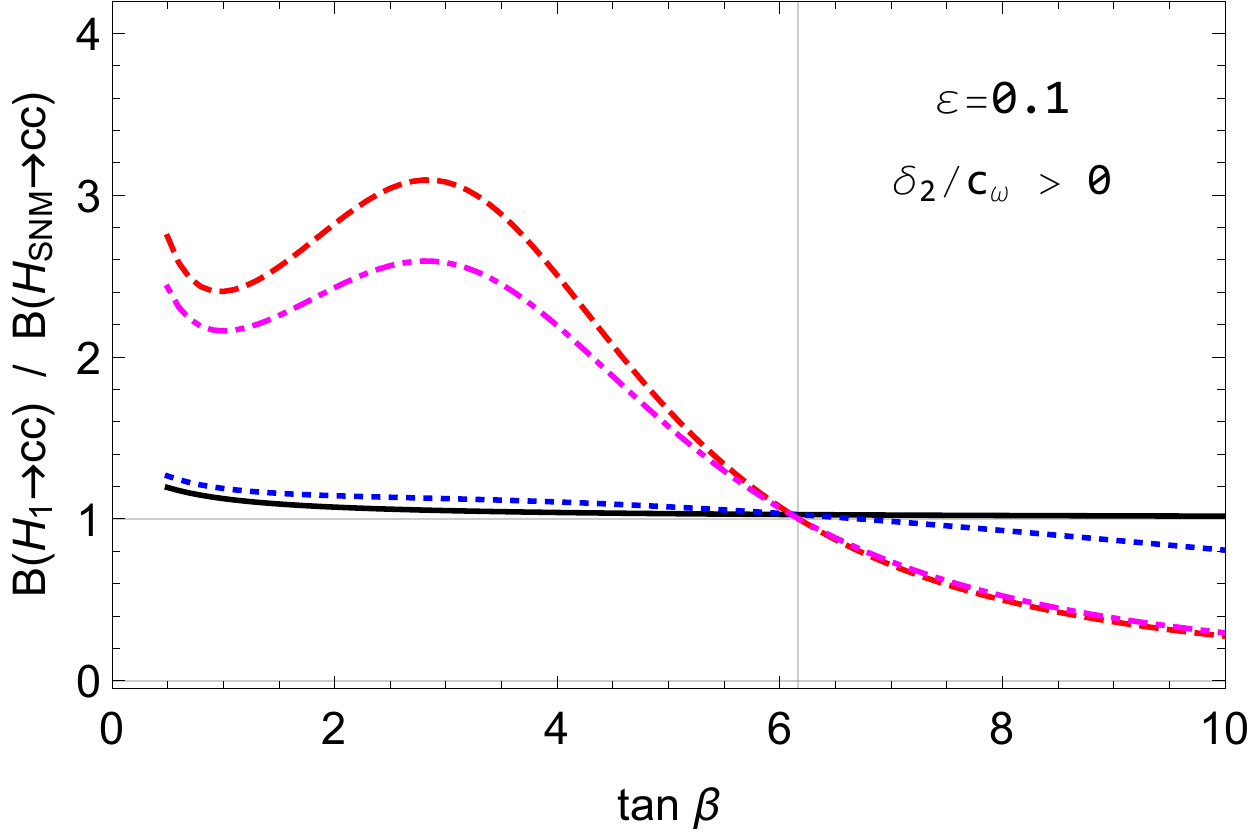}
\includegraphics[height=1.6in,width=2in]{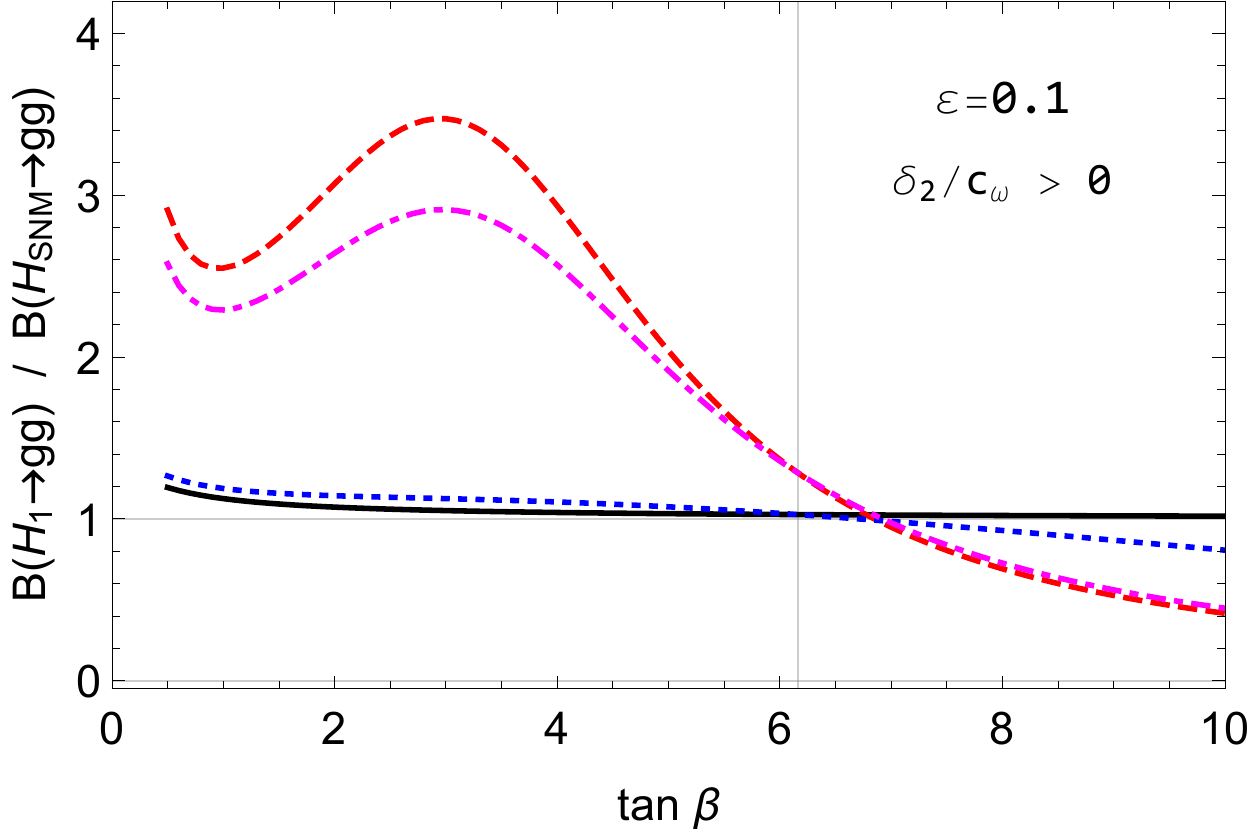}
\includegraphics[height=1.6in,width=2in]{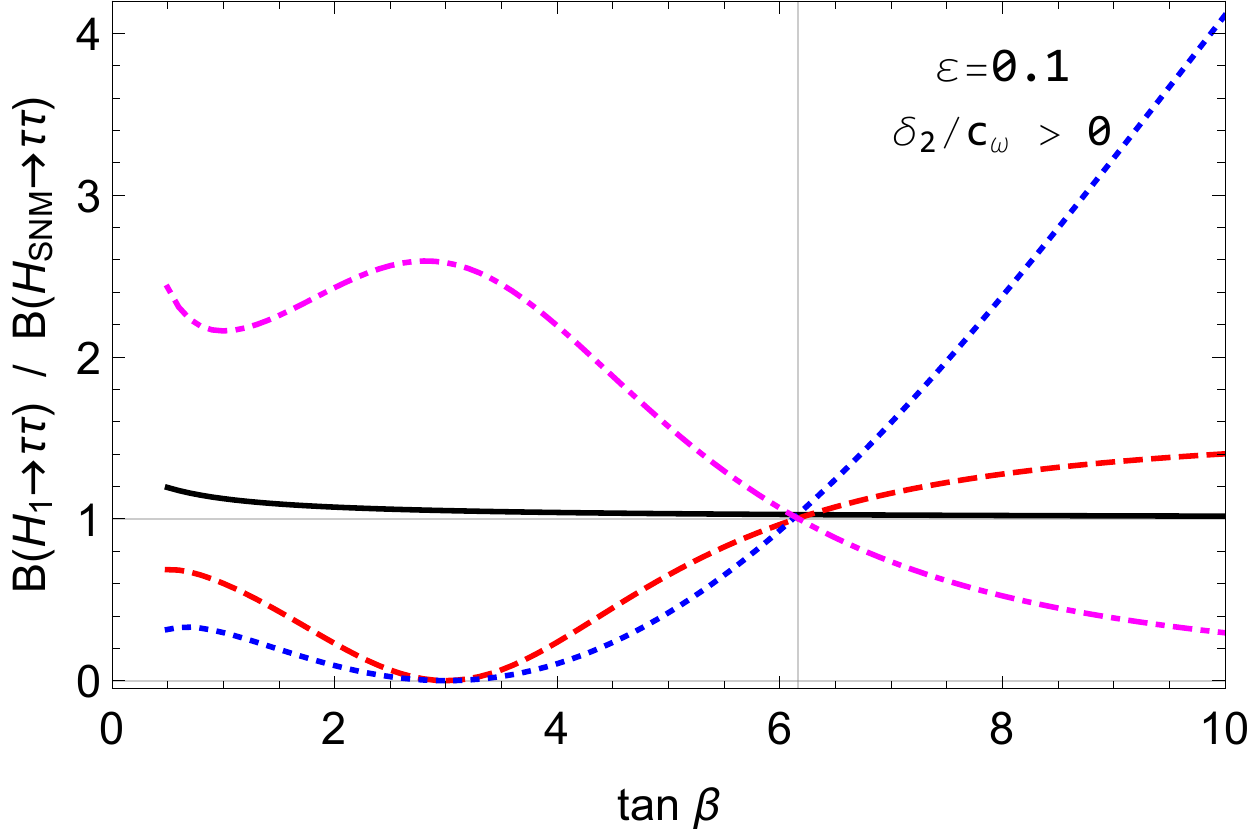}
\includegraphics[height=1.6in,width=2in]{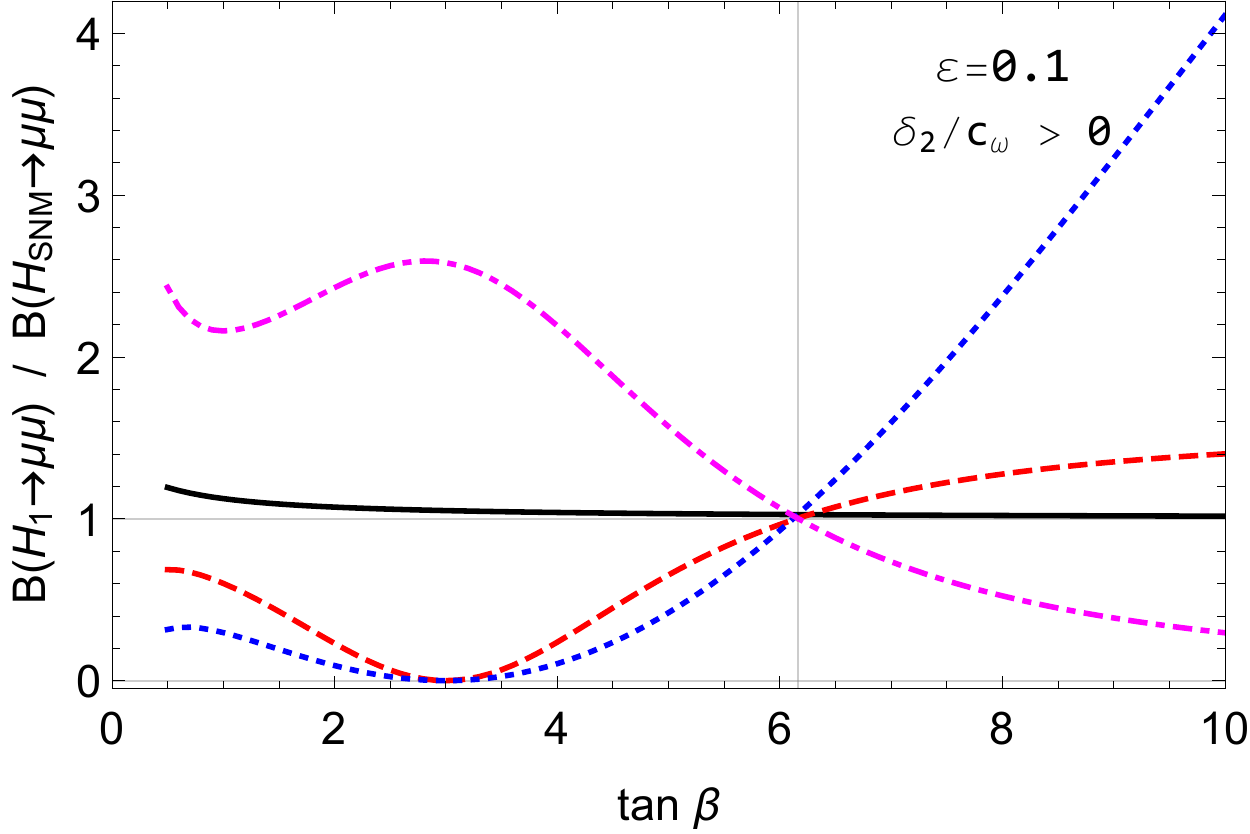}
\includegraphics[height=1.6in,width=2in]{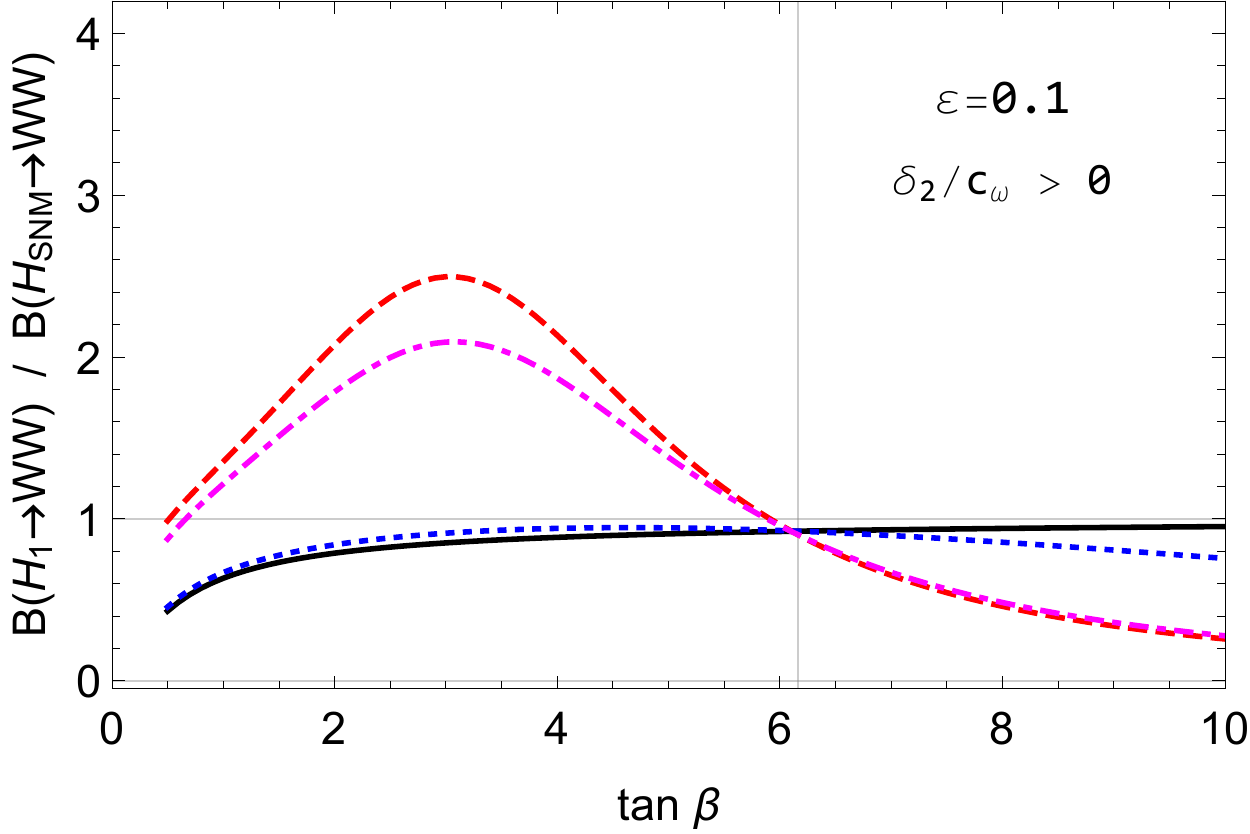}
\includegraphics[height=1.6in,width=2in]{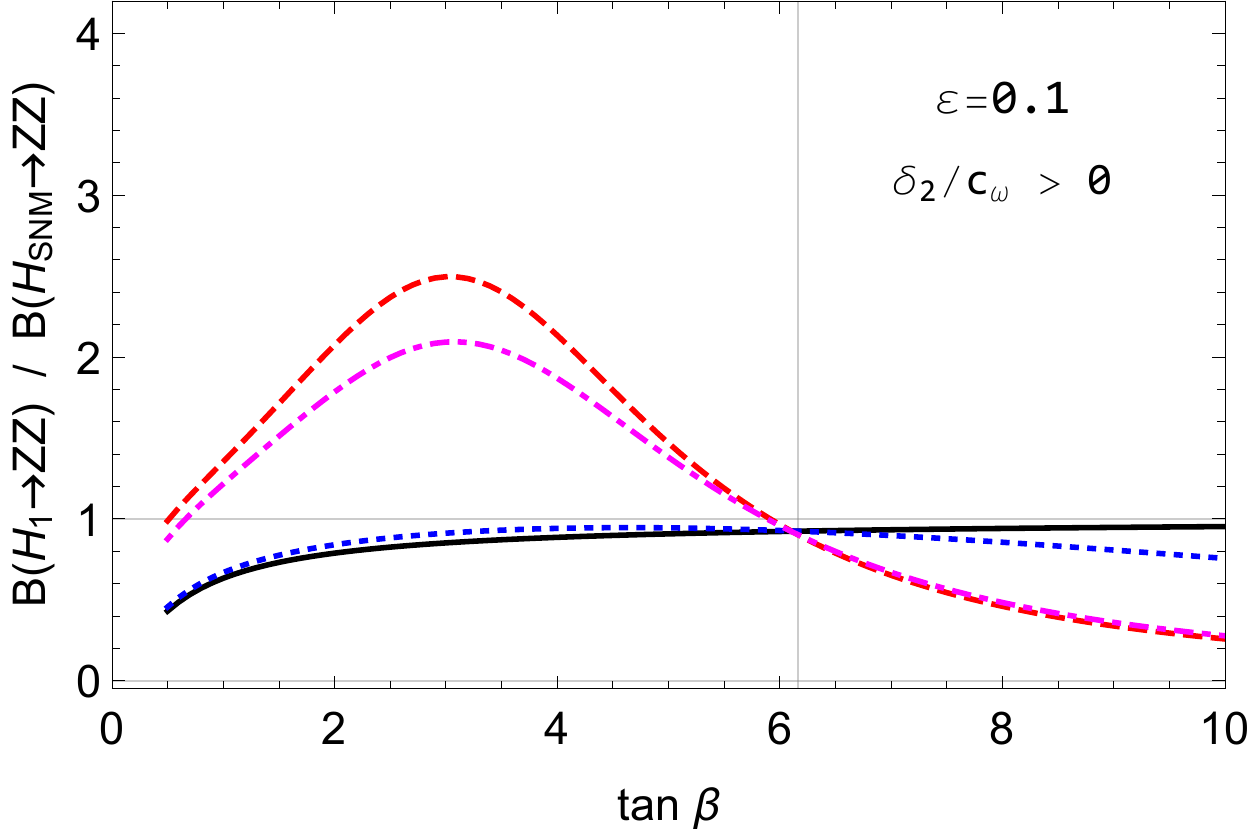}
\includegraphics[height=1.6in,width=2in]{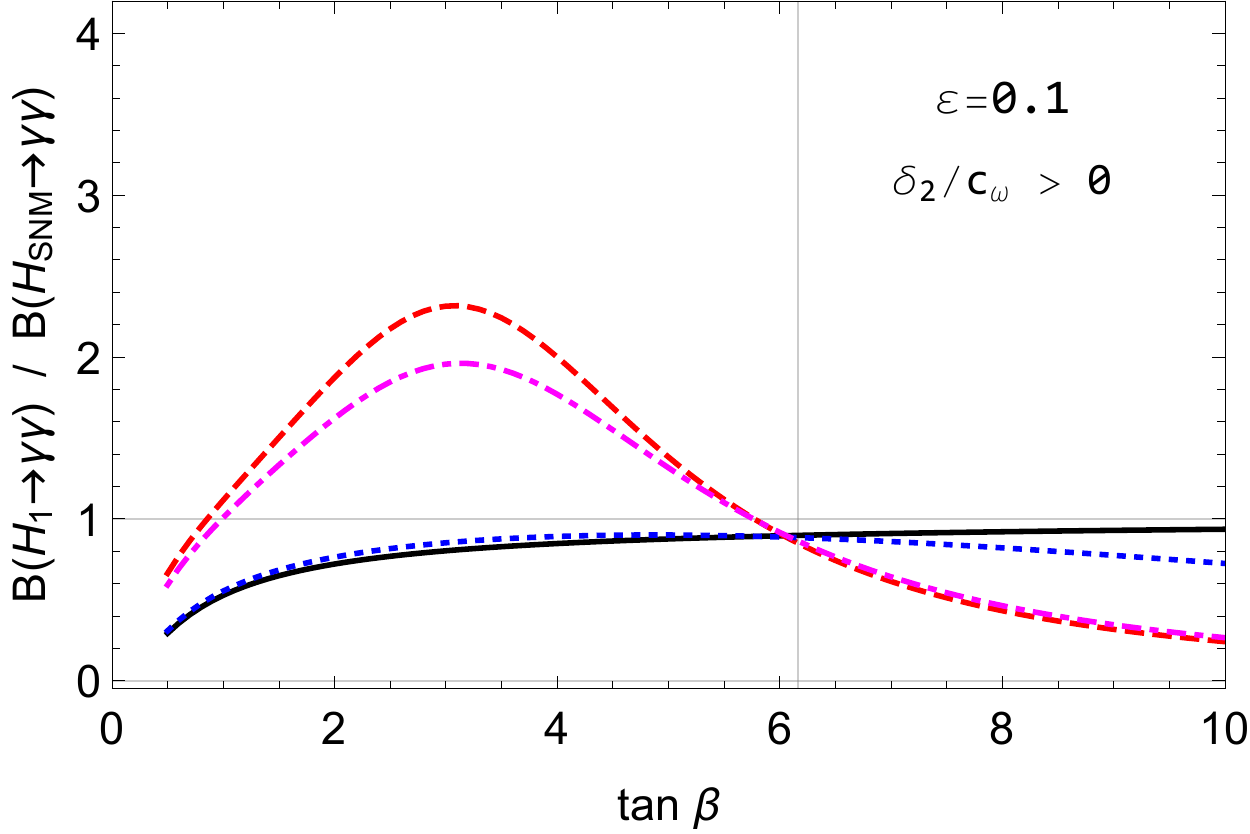}
\includegraphics[height=1.6in,width=2in]{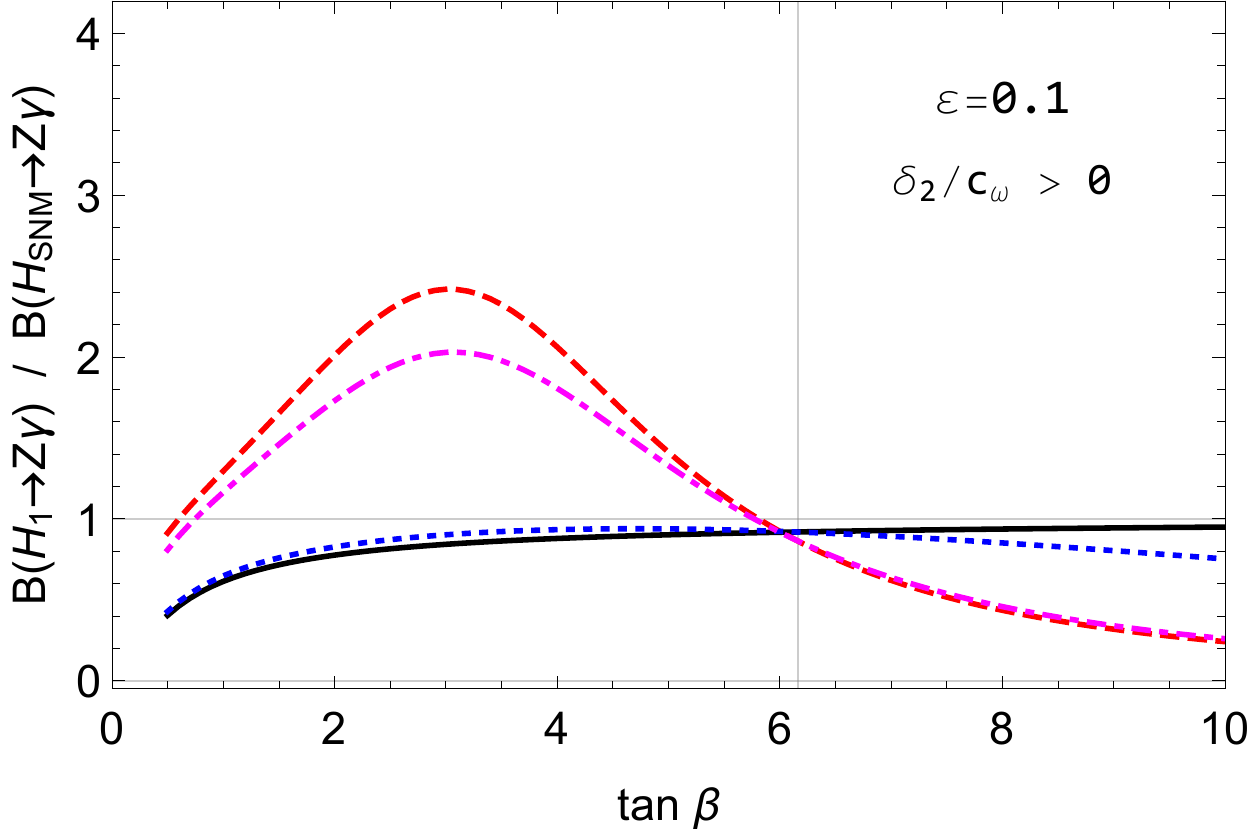}
\caption{\label{fig:h1br}
The ratios of branching ratios
$B(H_1\to {\cal D})/B(H_{\rm SM}\to {\cal D})$ as functions of $t_\beta$
in the four types of 2HDMs:
${\cal D}=b\bar b$ (upper left),
$c\bar c$ (upper middle),
$gg$ (upper right),
$\tau\tau$ (middle left),
$\mu\mu$ (middle middle),
$WW$ (middle right),
$ZZ$ (lower left),
$\gamma\gamma$ (lower middle), and
$Z\gamma$ (lower right).
The lines and the input parameters are the same as in Fig.~\ref{fig:h1gam}
}
\end{figure}
In Fig.~\ref{fig:h1gam}, we show the total decay width of
the lightest Higgs boson $\Gamma_{\rm tot}(H_1)$ normalized
to the SM one $\Gamma_{\rm tot}(H_{\rm SM})=4.122$ MeV. We consider
all four types of 2HDMs taking
$M_{H_1}=M_{H_{\rm SM}}=125.5$ GeV  and ${\rm sign}[\delta_2/c_\omega]=+1$ 
with $t_\beta$ varied.
For $\epsilon$,  we take $\epsilon/2=0.05$ adopting
the $1\sigma$ error of 5\% for the $g_{_{H_1VV}}=\sqrt{1-\epsilon}$ coupling 
obtained by the 4-parameter fit to the current LHC Higgs data~\cite{Cheung:2018ave}
\footnote{The varied 4 parameters are 
the three third-generation Yukawa couplings 
and the coupling to a pair of massive vector bosons:
$g^S_{H_1\bar t t}$, $g^S_{H_1\bar b b}$, $g^S_{H_1\tau\tau}$, and $g_{_{H_1VV}}$.}.
Taking into account that the error of the total decay width of the 125.5 GeV Higgs boson
is about 10\% at 95\% confidence level~\cite{Cheung:2018ave,ATLAS:2020cjb},
we find that the type-I 2HDM is reasonably viable for $t_\beta\gsim 2$ against
the LHC Higgs precision data.
The type-III 2HDM also seems viable unless $t_\beta$ is too large.
But we note this is only true when sizeable deviation of the branching ratios of 
the decay modes into charged leptons from their SM values is tolerated,
see the blue dotted lines in the middle-left and middle-middle panels in Fig.~\ref{fig:h1br}.
In the type-II and type-IV 2HDMs, the total decay width is consistent with 
the current LHC Higgs precision data only around $t_\beta=1/2$ and $6$. 
But, around $t_\beta=1/2$, we observe discrepancies in the decay modes into fermions and
gluons, see the dashed and dash-dotted lines in Fig.~\ref{fig:h1br}. Specifically, 
we find the reduction of the branching ratio
into $b\bar b$, which is mainly compensated by the increment in $H_1\to gg$,
and conclude that the first case with $t_\beta \sim 1/2$ should be valid 
only for the smaller values of $\epsilon$, 
see the horizontal lines locating the positions of
$\sqrt{1-\epsilon}\pm\sqrt\epsilon$  in Fig.~\ref{fig:gsh1ff}.
On the other hand, around $t_\beta=6$ where the absolute values of the 
Yukawa couplings are exactly aligned with the SM ones for 
the given value of $\epsilon=0.1$, 
the type-I and type-III 2HDMs are
consistent with the current LHC Higgs precision data.
In the wrong-sign alignment limit of the Yukawa couplings,
the type-II and -IV 2HDMs are also made feasible if 
the preference of the positive sign of the bottom-quark Yukawa coupling, 
which is currently observed at $\sim 1.5\sigma$ level,
has not been taken seriously.

\section{Conclusions}
We study the alignment of Yukawa couplings 
in the framework of general 2HDMs
identifying the lightest neutral Higgs boson as the 125 GeV one discovered at the LHC
while the two heavier neutral Higgs bosons are allowed to mix 
in the presence of nontrivial CP-violating phases in the Higgs potential.
We summarize our major findings as follows:
\begin{enumerate}
\item
In the type-I 2HDM,
the alignment of Yukawa couplings without decoupling
could be easily achieved as long as $\tan\beta \gsim 2$ even for 
the largest possible value of $\epsilon = 0.1$ allowed
by the current LHC Higgs precision data on the total decay width.
\item
Otherwise, we find that the Yukawa couplings decouple much slowly
compared to the $g_{_{H_1VV}}$ coupling.
For $\epsilon=0.01$, the $g_{_{H_1VV}}$ coupling deviates from its SM
value by only 0.5\% but 
the bottom-quark Yukawa coupling in type-II and -IV 2HDMs and 
the tau-lepton Yukawa coupling in type-II and -III 2HDMs  could
deviate from their SM values by more than 100\% when $t_\beta > 10$.
\item
In the wrong-sign alignment limit of the Yukawa couplings
where $t_\beta=(1+\sqrt{1-\epsilon})/\sqrt\epsilon$,
all four types of 2HDMs are viable against the LHC Higgs precision data
with the Yukawa couplings to the down-type quarks and/or
those to the charged leptons being equal in strength but
opposite in sign to the corresponding SM ones.
In contrast,
the magnitude and sign of the up-type quark Yukawa couplings
remain the same as in the SM.
\end{enumerate}

%
%
\section*{Acknowledgment}
This work was supported by the National Research Foundation (NRF) of Korea 
Grant No. NRF-2016R1E1A1A01943297 (J.S.L., J.P.)  and
No. NRF-2018R1D1A1B07051126 (J. P.).
The work of S.Y.C was
in part by Basic Science Research Program through the NRF of Korea
Grant No. NRF-2016R1D1A3B01010529 and in part by the CERN-Korea theory
collaboration.


\end{document}